\documentclass[%
aip,
amsmath,amssymb,
reprint,%
]{revtex4-1}

\usepackage{amsfonts}
\usepackage[english]{babel}
\usepackage{tumcolors}
\usepackage{tummath}
\usepackage{siunitx}
\usepackage{tikz}
\usepackage{svg}
\usepackage{pgfplots}
\usepackage{pgfplotstable}
\usepackage{environ}
\usepackage[hidelinks]{hyperref}
\usepackage[autostyle, english=american]{csquotes}
\usepackage{orcidlink}
\usepackage{qcircuit}
\usepackage{adjustbox}
\usepackage{upgreek}
\MakeOuterQuote{"}

\usepackage{multirow}
\usepackage{booktabs}
\usepackage{mathtools}
\usepackage{bm}
\usepackage{array}
\usepackage{listings}

\newcommand{\figref}[2][{}]{\hyperref[#2]{\figurename~\ref{#2}#1}}
\newcommand{\tabref}[2][{}]{\hyperref[#2]{\tablename~\ref{#2}#1}}
\renewcommand{\imu}{\i}

\DeclareSIUnit{\arbitraryunit}{arb.u.}
\DeclareSIUnit{\period}{period}
\DeclareSIUnit{\dBc}{dBc}

\usetikzlibrary{arrows, decorations.markings, external}
\usetikzlibrary{3d, positioning, fit, calc}
\usetikzlibrary{arrows.meta, shapes, decorations.pathmorphing}

\usepgfplotslibrary{units,fillbetween}
\pgfplotsset{compat=newest}

\pgfplotsset{
	layers/axis lines on top/.define layer set={
		axis grid,
		pre main,
		axis background,
		main,
		axis ticks,
		axis tick labels,
		axis lines,
		axis descriptions,
		axis foreground,
	}{/pgfplots/layers/standard},
}
\pgfplotsset{yaxis color style/.style={y axis line style = {#1},
		y tick label style= {#1},
		y tick style= {#1},
		ylabel style = {#1},
}}

\hypersetup{
	colorlinks=true,
	linkcolor=blue,
	citecolor=blue,
	filecolor=blue,
	urlcolor=blue
}
\begin{document}
	
	\preprint{AIP/123-QED}
	
	\title{Simulating spin biology using a digital quantum computer: Prospects on a near-term quantum hardware emulator}
	
	\author{Pedro H. Alvarez\orcidlink{0000-0003-3793-7545}}
	\affiliation{Institut für Physik, Carl von Ossietzky Universität, Oldenburg, Germany}
	\affiliation{Institut für Biologie und Umweltwissenschaften, Carl von Ossietzky Universität, Oldenburg, Germany}
	\affiliation{Department of Chemistry, University of Oxford, Oxford, UK}
	\affiliation{Department of Electrical and Computer Engineering, University of California, Los Angeles, CA, USA}
	\affiliation{Instituto de F\'isica Gleb Wataghin, Universidade de Campinas, Campinas, SP, Brazil}
	
	\author{Farhan T. Chowdhury\orcidlink{0000-0001-8229-2374}}
	\affiliation{Department of Electrical and Computer Engineering, University of California, Los Angeles, CA, USA}
	\affiliation{Department of Physics and Living Systems Institute, University of Exeter, Exeter, UK}
	
	\author{Luke D. Smith\orcidlink{0000-0002-6255-2252}}
	\affiliation{Department of Physics and Living Systems Institute, University of Exeter, Exeter, UK}
	
	\author{Trevor J. Brokowski}
	\affiliation{Department of Electrical and Computer Engineering, University of California, Los Angeles, CA, USA}
	\affiliation{Computational Biology and Biomedical Informatics (CBB), Yale University, West Haven, CT, USA}
	
	\author{Clarice D. Aiello\orcidlink{0000-0001-7150-8387}}
	\affiliation{Department of Electrical and Computer Engineering, University of California, Los Angeles, CA, USA}
	
	\author{Daniel R. Kattnig\orcidlink{0000-0003-4236-2627}}
	\affiliation{Department of Physics and Living Systems Institute, University of Exeter, Exeter, UK}
	
	\author{Marcos C. de Oliveira\orcidlink{0000-0003-1745-4888}}
	\affiliation{Instituto de F\'isica Gleb Wataghin, Universidade de Campinas, Campinas, SP, Brazil}
	
	\date{\today}
	\begin{abstract}
		Understanding the intricate quantum spin dynamics of radical pair reactions is crucial for unraveling the underlying nature of chemical processes across diverse scientific domains. In this work, we leverage Trotterization to map coherent radical pair spin dynamics onto a digital gate-based quantum simulation. Our results demonstrated agreement between the idealized noiseless quantum circuit simulation and established master equation approaches for homogeneous radical pair recombination, identifying approximately 15 Trotter steps to be sufficient for faithfully reproducing the coupled spin dynamics of a prototypical system. By utilizing this computational technique to study the dynamics of spin systems of biological relevance, our findings underscore the potential of digital quantum simulation (DQS) of complex radical pair reactions and builds the groundwork towards more utilitarian investigations into their intricate reaction dynamics. We further investigate the effect of realistic error models on our DQS approach, and provide an upper limit for the number of Trotter steps that can currently be applied in the absence of error mitigation techniques before losing simulation accuracy to deleterious noise effects.  
	\end{abstract}
	
	\maketitle
	\section{\label{sec-introduction}Introduction}  Quantum devices and algorithms have the potential to simulate chemical systems in an efficient manner \cite{Bauer2020}, with the promise of exceeding the capabilities of state-of-the-art methods for classical devices \cite{guzik2024, li2021quantum, nitro17, trev_ieee} in the long term. Computational challenges arising out of the need for simulating spin-chemical systems of realistic complexity in the emerging field of quantum biology \cite{kattnig2023nat, Zadeh_Haghighi_2022, Hore2020, quantum3010006, cai2017} thus presents a prime area in which to explore digital quantum simulation (DQS) and its applications to molecular and biological systems \cite{reiher23}. Many migratory animal species, such as some night-migratory songbirds, have the ability to navigate aided by sensing the Earth's geomagnetic field \cite{cai2024, ieee09, ieee08}. Correlative evidence suggests that a photo-initiated electron transfer reaction in the flavoprotein cryptochrome, resulting in a radical pair, may underlie this magnetic field sensitivity via the radical pair mechanism \cite{Wiltschko_2009, Hore2016, kattnig17}. 
	
	Upon photo-excitation, radical pair electron spins are in an initially entangled state, namely the singlet state, which is sensitive to external magnetic fields via the Zeeman interaction and the surrounding nuclear spins via hyperfine interactions. As a result of these interactions, which are locally different, each electron precesses at a different rate resulting in a coherent interconversion between singlet and triplet states, dependent on the orientation to the external field. After some time the radical pair forms a singlet or triplet product. These products have different chemical reactivity and the amount of each chemical reaction product leads to correspondingly differing physiological properties, which thus provide directional information about the external field \cite{kt19, cai2024} downstream. The angular information from the geomagnetic field combined with navigational information from solar cues, the night sky, and landmarks, bestow migratory birds with a precise compass sense for navigation. While classical quantum emulation tools have enabled simulations of radical pair reactions that account for the signature traits of magnetic field sensing observed experimentally \cite{kattnig17, PhysRevE.92.012720, kt19, Ikeya_2021, Woodward2022}, they are limited in terms of accounting for the number of hyperfine interactions that can be treated, ultimately falling short of the total number present in a biological setting \cite{smith2024optimality, chow2023, ieee18}. 
	
	As the number of elements involved in a radical pair reaction increases linearly, the computational time and complexity increases exponentially, due to the corresponding growth in Hilbert-space dimension. This motivates us to leverage the close analogy between qubits and spin-$\frac{1}{2}$ particles of radical pairs to simulate the relevant spin dynamics of radical pair systems on a quantum device with the aim of proposing a method capable of tackling larger and more physically relevant biological spin systems in reasonable computation time. Although other quantum biological systems were broadly explored with DQS schemes \cite{guzik2024, nitro17, li2021quantum, Guimaraes2020, prot23, oh2023}, simulating radical pair dynamics on quantum computers is a relatively new endeavor. 
	
	Our work assesses the viability of Trotterization for simulating radical-pair spin dynamics. Although widely established as a tried and tested method in DQS, it has not been considered so far in a quantum simulation of spin biological effects in radical-pairs relevant to magnetoreception. We address this by leveraging Trotterization to simulate the unitary evolution of a radical pair spin system for symmetric radical pair recombination, for which the non-unitary evolution due to recombination can be accounted for in post via an exponential population decay. The method shows promise in replicating qualitative features of the spin dynamics, in particular allowing for the simulation of the time-integrated reaction yields, the actual experimental observable providing signatures of an underlying spin-mediated mechanism, under different orientations to the magnetic field.
	
	\section{\label{sec:model}Methods \& Model}
	Photo-reduction in cryptochrome implicates a radical pair comprising the reduced flavin adenine dinucleotide (FAD) co-factor and an oxidized surface-expose  tryptophan (TrpH) \cite{Hore2016}. Photon absorption by FAD leads to photo-excitation, permitting it to undergo an electron transfer reaction. As a result an electron is transferred from the TrpH to form the radical pair, and as a result of hyperfine interaction between the electrons and the molecular nuclei, the spin state of the electrons coherently oscillates between the singlet and triplet states. This can be subject to modulation by external magnetic fields, and is sensitive to fields as weak as the geomagnetic field ($\sim 50\mu T$), with the overall recombination yield in the singlet and triplet populations dependent on the angle between the radical pair and the field lines. We can define a Hamiltonian that contains the internal interactions of the spins and models the state exchange during the singlet-triplet interconversion prior to recombination as
	\begin{equation}\label{eq:hamiltonian1}
		\begin{split}
			\hat{H} = g_1\mu_B \bold{B}\cdot\hat{\bold{S}}_1 +g_2\mu_B \bold{B}\cdot\hat{\bold{S}}_2 + \\ \sum_{k=1}^{N_1} \hat{\bold{I}}_{1,l} \cdot \bold{A}_{1,l} \cdot \hat{\bold{S}}_1 + \sum_{k=1}^{N_2} \hat{\bold{I}}_{2,k} \cdot \bold{A}_{2,l} \cdot \hat{\bold{S}}_2,
		\end{split}
	\end{equation}
	which contains the Zeeman interaction between the field and the electron spins and the hyperfine interaction between the molecular nuclei and the electron spin\cite{PhysRevE.92.012720}. These are a minimal set of interactions necessary to allow for magnetosensitivity, but there are more complex models including, for example, dipolar interaction between electron spins\cite{Fay2020}. The Zeeman effect is the split in the energy levels caused by a magnetic field\cite{ZEEMAN}. In Eq. \ref{eq:hamiltonian1} the Zeeman interaction between the spin of electron $i$ and the field $\bold{B}$ is given by
	\begin{equation}
		\hat{H}_z = g_i\mu_B \bold{B}\cdot\hat{\bold{S}}_i,
	\end{equation}
	where $g_i$ is the g-factor of electron $i$, $\mu_B$ is the Bohr magneton and $\hat{\bold{S}}_i$ is the spin operator for electron $i$. In weak magnetic fields, such as the geomagnetic field, the Zeeman interaction does not cause significant singlet-triplet interconversion. For the field direction to be detectable there must be an interaction that changes the spin state from singlet to triplet, which can be accounted by the 
	hyperfine interaction between the spin of the molecular nuclei and the electron spins, defined as\begin{equation}
		\hat{H}_{hf} = \sum_{l=1}^{N_i} \hat{\bold{I}}_{i,l} \cdot \bold{A}_{i,l} \cdot \hat{\bold{S}}_i,
	\end{equation}
	with $N_i$ being the number of interacting molecular nuclei, $\hat{\bold{I}}_{i,l}$ denoting the spin operator of nucleus $l$ interacting with electron $i$ and $\bold{A}_{i,l}$ being the hyperfine tensor that defines the interaction.

\subsection{Prototypical radical pair model}
We consider a simplified radical pair model, commonly used in theoretical assessments of the radical pair mechanism of magnetoreception \cite{PhysRevE.92.012720, Vedral2011}, comprising of two electron spins with only one hyperfine-coupled proton, described by a Hamiltonian of the form
\begin{equation}
	\hat{H} = g_1\mu_B \bold{B}\cdot\hat{\bold{S}}_1 +g_2\mu_B \bold{B}\cdot\hat{\bold{S}}_2 + \hat{\bold{I}}_{1,1} \cdot \bold{\bold{A}}_{1,1} \cdot \hat{\bold{S}}_1, \label{eq:Hamiltonian}
\end{equation}
where $\hat{\mathbf{S}}_{i}$ denotes the electron spin $i$, $\hat{\mathbf{I}}_{1,1}$ denotes the nuclear spin, $\mathbf{A}_{1,1}$ denotes the hyperfine tensor coupling the nuclear spin to the first electron spin and $\bold{B} = B(\sin{\theta}\cos{\phi}, \sin{\theta}\sin{\phi}, \cos{\theta})$ denotes the external magnetic field, where angles $(\theta, \phi)$ represent the direction of the field in relation to the radical pair's fixed axis\cite{PhysRevE.92.012720}. Furthermore, as in \cite{Vedral2011}, we assume an axially symmetric anisotropic hyperfine coupling tensor, such that the components of $\mathbf{A}_{1,1}$ are $A_{x}=A_{y}=A_{z}/2$. This choice of hyperfine coupling imitates the dominant hyperfine interaction of the radical pair within cryptochrome. 

Whilst the Hamiltonian in Eq.\ \ref{eq:Hamiltonian} models a physical abstraction of a realistic radical pair, which in reality contain numerous hyperfine interactions and inter-radical interactions \cite{Fay2020}, it provides a basic model capable of accounting for the essential characteristic of a directional magnetic field effect on singlet-triplet interconversion and offers an entry point for comparison to simulations for similar models performed using established classical techniques, upon which more realistically complex simulations can be built. Additionally, the axially symmetric hyperfine coupling tensor allows us to simplify our analysis by assuming $\phi = 0$\cite{PhysRevE.92.012720}. In accordance to the standard photo-initiated radical pair mechanism, we assume that the electron spins are initialised in the singlet state ($\ket{S} = (\ket{\uparrow\downarrow} - \ket{\downarrow\uparrow})/\sqrt{2}$). 
The dynamics of the system are conventionally modelled using the master equation formalism developed by Haberkorn\cite{Haberkorn1976}
\begin{equation}\label{haberkorn}
	\frac{d}{dt}\hat{\rho}(t) = -\frac{i}{\hbar}[\hat{H}, \hat{\rho}(t)] -\sum_{n = \{S,T\}}\{\frac{k_n}{2}\hat{P}_n, \hat{\rho}(t)\},
\end{equation}
where the first term accounts for the unitary dynamics. while the non-unitary reaction kinetics are incorporated via the second term, for which  $\hat{P}_n$ are the projectors onto the singlet ($n=S$) and triplet ($n=T$) states, and $k_n$ are the respective rates of recombination into singlet or triplet products. The singlet and triplet states form a complete basis, meaning that $\hat{P}_S + \hat{P}_T = \mathbb{I}$, therefore by obtaining the singlet population alone provides complete access to the population dynamics. The recombination rates are the inverse of the reaction's lifetime, so for a radical of $\sim 1\mu s$ lifetime, a rate of $1$MHz. For a prototypical radical pair in cryptochrome, rates on the order of $1$ MHz have been found\cite{Hore2020, Hore2016}. This choice ensures sufficient time for singlet-triplet interconversion to occur, yet maintains a lifetime still short enough such that spin relaxation processes do not dominate and negate magnetic field sensitivity.  

The Haberkorn equation\cite{Fay2018} does not preserve the trace of the density operator $\hat{\rho}$, therefore $\hat{\rho}$ does not represent a quantum state and the equation is not a Lindblad type equation in the traditional sense. Instead, the Haberkorn model functions more as a rate equation, where $\hat{\rho}$ represents an ensemble of states undergoing chemical reactions. The reaction kinetics, described by projection operators account for the loss of population due to recombination and product formation events. This kind of formalism is commonly used in theoretical spin chemistry to simulate the dynamics of chemical reactions and incorporate both the elements of coherent spin dynamics and the incoherent reaction kinetics. In the case of symmetric decay e.g. $k_S = k_T = k$ it is possible to solve for the unitary part of the equation\cite{Bargon2006, Lawler1971, Brocklehurst1996}
\begin{equation}
	\frac{d}{dt}\hat{\rho'}(t) = -\frac{i}{\hbar}[\hat{H}, \hat{\rho'}(t)]
\end{equation}
and then incorporate reaction kinetics by multiplying the resulting density operator by an exponential decay $\hat{\rho}(t) = e^{-kt}\hat{\rho'}(t)$.

Whilst basic spin-chemical models can be solved numerically using existing classical techniques, the dimensionality of the system increases exponentially with the number of hyperfine coupled nuclear spins included, making it intractable to solve bigger systems with existing methods. In this work we assess this simple model to understand the impact of current hardware noise on our Trotterization approach and gain insight into its capability for solving larger systems via DQS. The parameters relevant to our simulations are provided in Table \ref{table2} in Appendix \ref{parameters}.

\subsection{Trotter Simulation}
%

We utilize the Trotter-Suzuki expansion \cite{Suzuki1976} to decompose a time evolution operator, defined by the Hamiltonian given by Eq.\ \ref{eq:Hamiltonian}, into a quantum circuit capable of simulating unitary evolution. The method emerges from the limit
\begin{equation}
	e^{-\frac{it}{\hbar} \hat{H}} = e^{-\frac{it}{\hbar}\sum_{i=1}^N \hat{H}_i} = \lim_{n\longrightarrow \infty}\left( \Pi_{i=0}^N e^{(-\frac{it}{\hbar}\frac{\hat{H}_i}{n})}\right)^n,
\end{equation}
where $\hat{H}_i$ is an additive component of the Hamiltonian, and $\hat{H}_i \propto \hat{\sigma_j}\otimes\hat{\sigma_k}\otimes\hat{\sigma_l}$ where $\hat{\sigma}_{j}$, $\hat{\sigma}_{k}$, and $\hat{\sigma}_{l}$ can be any of the Pauli matrices $X, Y, Z$ or the identity matrix $I$. If the order of Trotterization $n$ (number of steps) is sufficiently large, we have a good approximation of the whole unitary evolution. This is computationally simple, and many quantum computing programming languages already have algorithms implemented to decompose a Hamiltonian using this scheme. Trotterization is particularly advantageous for systems with a short lifetime, such as radical pairs, which have a lifetime in the order of $\mu s$, since the Trotterized dynamics are expected to diverge from the true dynamics after some time, as determined by the order of Trotterization (cf.\ Fig. \ref{fig_noisy} in Appendix \ref{trotter_noise}).
\subsection{Quantum Circuit}
\begin{figure}[h]
	\centering
	\begin{equation*}
		\Qcircuit @C=0.7em @R=2em {
			\ket{e_1} & & & \gate{X} & \gate{H} & \ctrl{1} & \qw & \multigate{2}{e^\frac{-it\hat{H}}{n\hbar}} & \multigate{2}{e^\frac{-it\hat{H}}{n\hbar}} &&& \qw & \ctrl{1} & \gate{H} & \meter\\
			\ket{e_2} & & & \gate{X} & \qw & \targ & \qw &  \ghost{e^\frac{-it\hat{H}}{n\hbar}}&  \ghost{e^\frac{-it\hat{H}}{n\hbar}} & ... && \qw & \targ & \qw & \meter\\
			\ket{n_{1}} & & & \qw & \qw & \qw &\qw & \ghost{e^\frac{-it\hat{H}}{n\hbar}} & \ghost{e^\frac{-it\hat{H}}{n\hbar}} &&& \qw & \qw &\qw &\qw 
		}
	\end{equation*}
	\caption{Quantum circuit detailing the simulation process. The two $\ket{e_1}$ and $\ket{e_2}$ qubits represent the electron spin states and the $\ket{n_{1}}$ qubit represents the nuclear spin states. The electron qubits are prepared in the singlet states, then the Trotterized gates evolve the system unitarily up to a time $t$ under the Hamiltonian $\hat{H}$ and with order of Trotterization $n$. After the evolution, the states of the electron qubits are prepared such that the measurement result $\ket{11}$ represents the singlet state.}
	\label{circuit}
\end{figure}
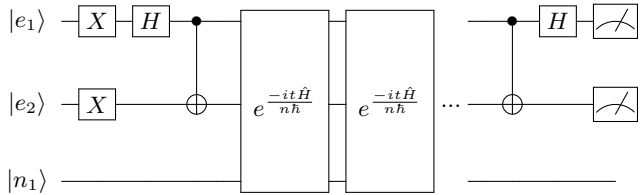

Since qubits, as two level spin systems, are directly analogous to spin-$\frac{1}{2}$ particles, our circuit, implementing the Hamiltonian in Eq.\ \ref{eq:Hamiltonian} according to the Trotter-Suzuki expansion, requires three qubits. To prepare the electrons in the singlet state we initialize the circuit with gates that entangle the qubits. We assume that the qubits start at the $\ket{0}$ ($\ket{\uparrow}$) state, and then apply one $X$ gate to each electron qubit bringing them to $\ket{1}$ ($\ket{\downarrow}$), then a Hadamard and a CNOT gate are applied to the qubits to bring them to the singlet state (($\ket{01}-\ket{10})/\sqrt{2}$). This is followed by a series of gates representing the unitary evolution of the Hamiltonian decomposed using the Trotter-Suzuki expansion, $e^\frac{-it{\hat{H}}}{n \hbar}$, that are repeated $n$ times, where $n$ is the order of Trotterization. Following this, a measurement is performed on the qubits to obtain the singlet state population. This process can be visualized by the quantum circuit in Fig. \ref{circuit}.
When testing the circuit in a quantum simulator, projective measurements are performed in the singlet-triplet basis. However, for the quantum device we perform a change of basis to $\ket{0}$ - $\ket{1}$, as it provides a more general way of expressing the circuit for any quantum architecture.
%
As the singlet and triplet states of the electronic spins are the relevant states for eliciting magnetic field sensitivity, the qubits representing nuclei are not considered during measurement. We assume that the initial nuclear spin state is the maximally mixed state such that ${\hat{\rho}(0) = \frac{1}{2} (\vert S, \downarrow \rangle \langle S, \downarrow\vert + \vert S, \uparrow \rangle \langle S, \uparrow\vert })$. To account for this in our quantum simulation, we run the circuit twice, sampling over the two possible nuclear spin state configurations ($\ket{\uparrow}$ and $\ket{\downarrow}$) and taking the average. Alternatively, the maximally mixed state of the nuclear subspace can be achieved through quantum state purification \cite{Wilde_2017}, avoiding multiple simulations of various pure states at the cost of additional auxiliary qubits.
%
\subsection{Measuring reaction yields}
Using the quantum circuit defined, we prepare the singlet state, representative of the initial state of the radical pair,  we apply Trotterized gates to emulate the unitary evolution of the Hamiltonian comprising Zeeman and Hyperfine interactions, and then perform a measurement on the electron qubits to obtain the singlet state population. Those populations are calculated as the ratio of measurements that resulted in $\ket{11}$ versus the other possibilities. With this in hand, and assuming symmetric recombination with rate constants $k_{S} = k_{T}= k$ we calculate the singlet yield
\begin{equation}
	\Phi_S = k \int_0^\infty \mbox{Tr}[\hat{P}_S \hat{\rho}(t)] \, dt,
\end{equation}
where $\hat{P_{S}}$ corresponds to the singlet projection operator and $\hat{\rho}(t)$ is the density operator of the electronic spins. The singlet yield is the fraction of the total product that recombined into singlet products, as opposed to triplet products\cite{Steiner1989}. Using the singlet yields, a widely used anisotropy measure of magnetic field sensitivity is defined as\cite{Wong2021}
\begin{align}
	\Delta_{S} = \Phi_{S_{max}} - \Phi_{S_{min}},
\end{align}
where the difference of the extremal singlet yields is taken over the orientations to the magnetic field.
%
\section{Results}
\begin{figure}[t]
	\includegraphics[width=1.02\columnwidth]{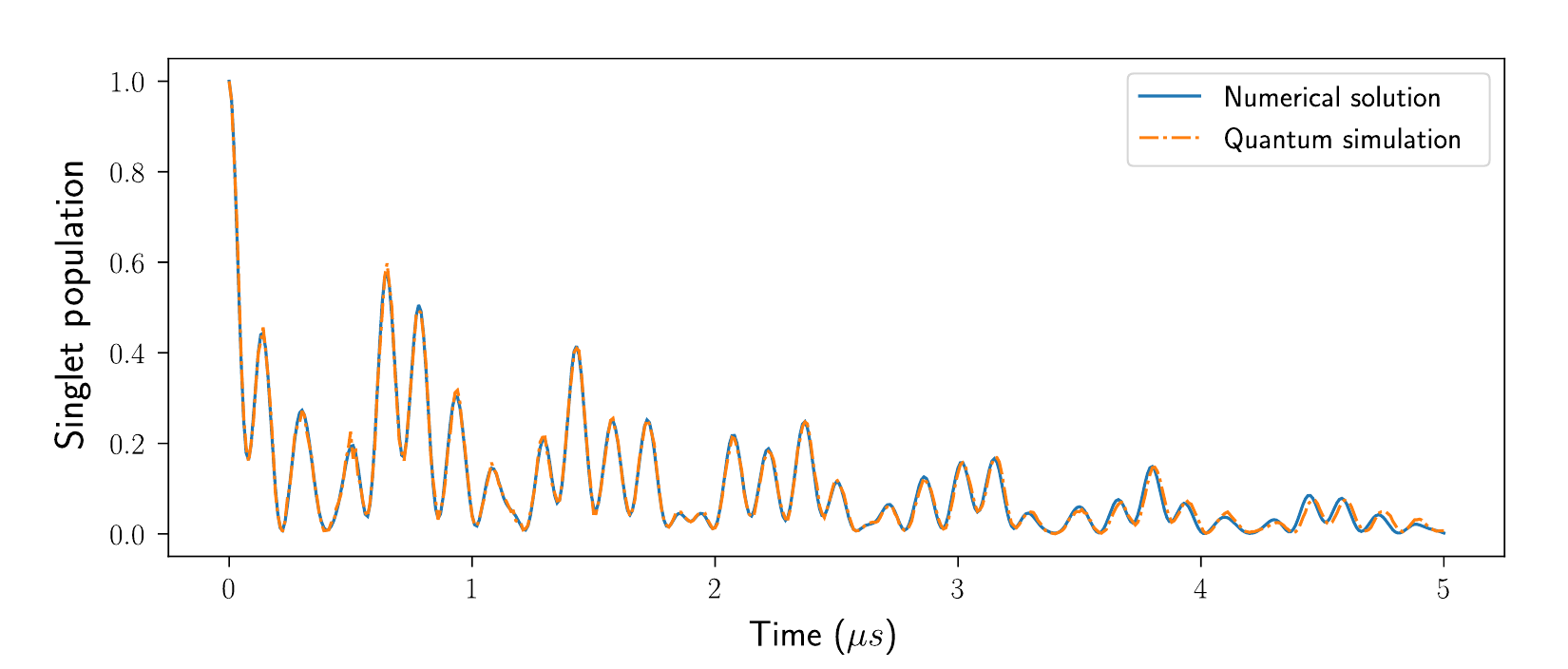}
	\caption{For $\theta = \pi/2$ we compare population results from quantum simulation to the numerical solutions of the differential equations. Multiplying the quantum simulation results by an exponential decay $e^{-kt}$ with the same rate $k$ as the model's recombination rate, the results from the numerical simulation are reproduced.}
	\label{fig_comparison}
\end{figure}

We first utilize an idealized quantum simulator to test our quantum circuit in the noiseless scenario. Using this, we directly extract the singlet population from the circuit in one shot, by calculating $\mbox{Tr}[\hat{P}_S \hat{\rho}(t)]$. As the simulator represents an ideal noiseless quantum computer, we can use a large Trotterization order, specifically $n=1024$, as seen in Fig.\ \ref{fig_comparison}. The circuit is used to simulate the unitary evolution of the system; the effects of recombination, which induce a non-unitary population decay, are incorporated by weighting the singlet population by the exponential decay $e^{-kt}$, as explained in the previous section. In the context of results calculated using the quantum circuit, we here refer to this process of incorporating the recombination kinetics as \textit{postprocessing}.

\subsection{Trotterization and noise}
\begin{figure}
	\centering
	\includegraphics[width=1.02\columnwidth]{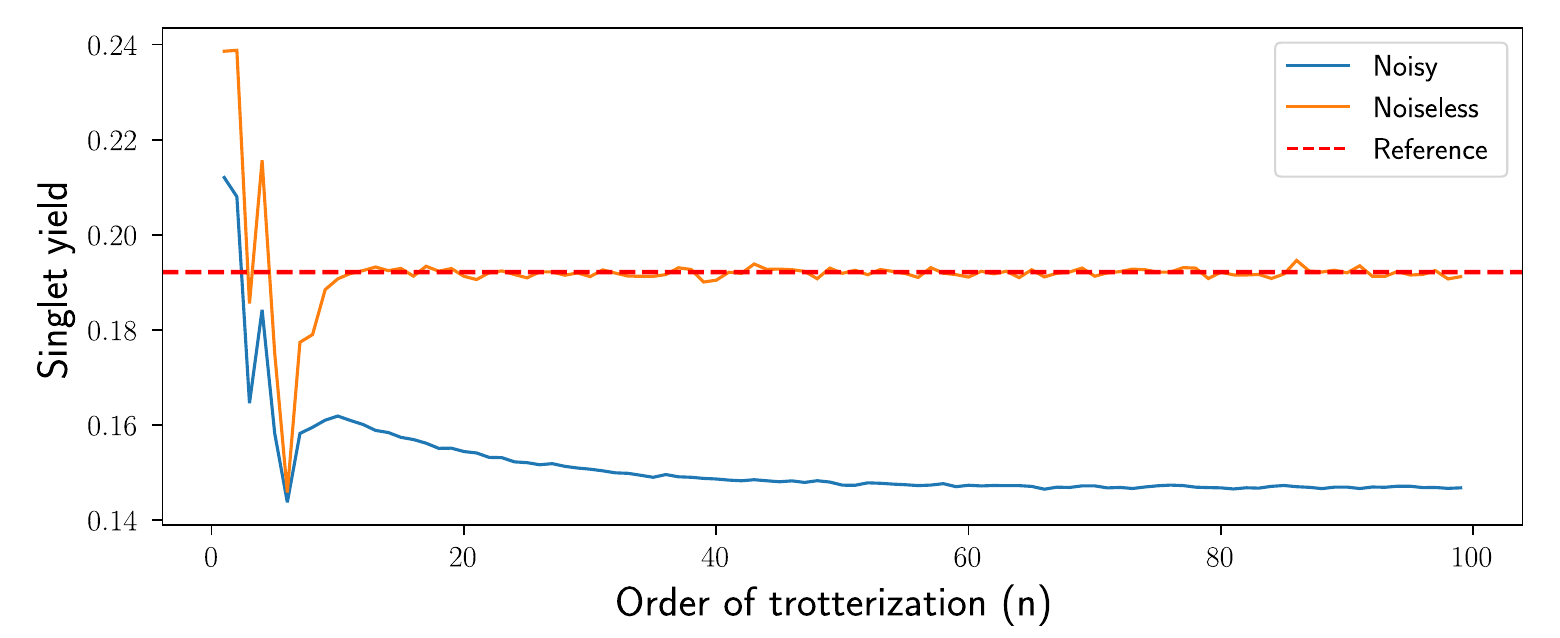}
	\caption{Singlet yield for $\theta = \pi/2$. The reference is the numerical solution for the singlet yield. Except for some amplitude loss, the noisy curve largely shows a similar behaviour to the noiseless until about $n=15$, when the influence of noise causes it to converge on a lower singlet yield than the noiseless simulation and the numerical solution.}
	\label{fig_trotter_order}
\end{figure}

With agreement found using a quantum simulator in the idealized noiseless case, we turn our attention to simulation of the circuit on IBM's noisy simulator with a noise profile mimicking that of IBM quantum devices. We utilize a noisy simulator to understand how our approach would behave in a realistic noisy intermediate scale (NISQ) quantum computer, which is affected by multiple sources of noise from the environment and is prone to errors during computations \cite{nisq18}. This allows us to test if our approach can still provide useful results when being executed in realistic quantum hardware, and to better understand how noise in quantum devices would influence the emulation of singlet yield and anisotropy in a gate-based DQS scheme. The noisy simulator was defined using the noise profile from the \textsf{ibmq\_jakarta} quantum device. To reduce the impact of noise, and maintain coherent singlet-triplet interconversion from the Zeeman and hyperfine interactions, we had to minimize the number of Trotterization steps $n$. In the noiseless simulation we used $n=1024$, but the number of gates in the circuit scales linearly with $n$\cite{Alvarez2023}. Such a high order of Trotterization adds about 13 thousand gates to the circuit, leaving it more vulnerable to noise. In Fig.\ \ref{fig_trotter_order}, we demonstrate that $n\approx15$ is a sufficient order of Trotterization to ensure accurate results in the noiseless simulation, beyond this point it is evident that increasing $n$ is not beneficial, but introduces additional error in the noisy simulation. In Appendix \ref{trotter_noise}, Fig. \ref{fig_steps_noisy} illustrates the impact of a higher order of Trotterization in more detail.
In the appendix, results from Fig. \ref{fig_noisy} show that our method can simulate the qualitative features of the unitary evolution in noisy quantum computers if the number of Trotterization steps is sufficiently reduced. However, to obtain a more quantitative agreement, one could harness efficient error mitigation strategies to improve the results, as well as capitalizing on the recent developments of more robust qubit architectures paving the way for eventual realisations of DQS in the fault tolerant regime \cite{lukinQCh, Bluvstein2024Logical, new24}. 

\subsection{Singlet yield}
\begin{figure} [h]
	\includegraphics[width=0.95\columnwidth]{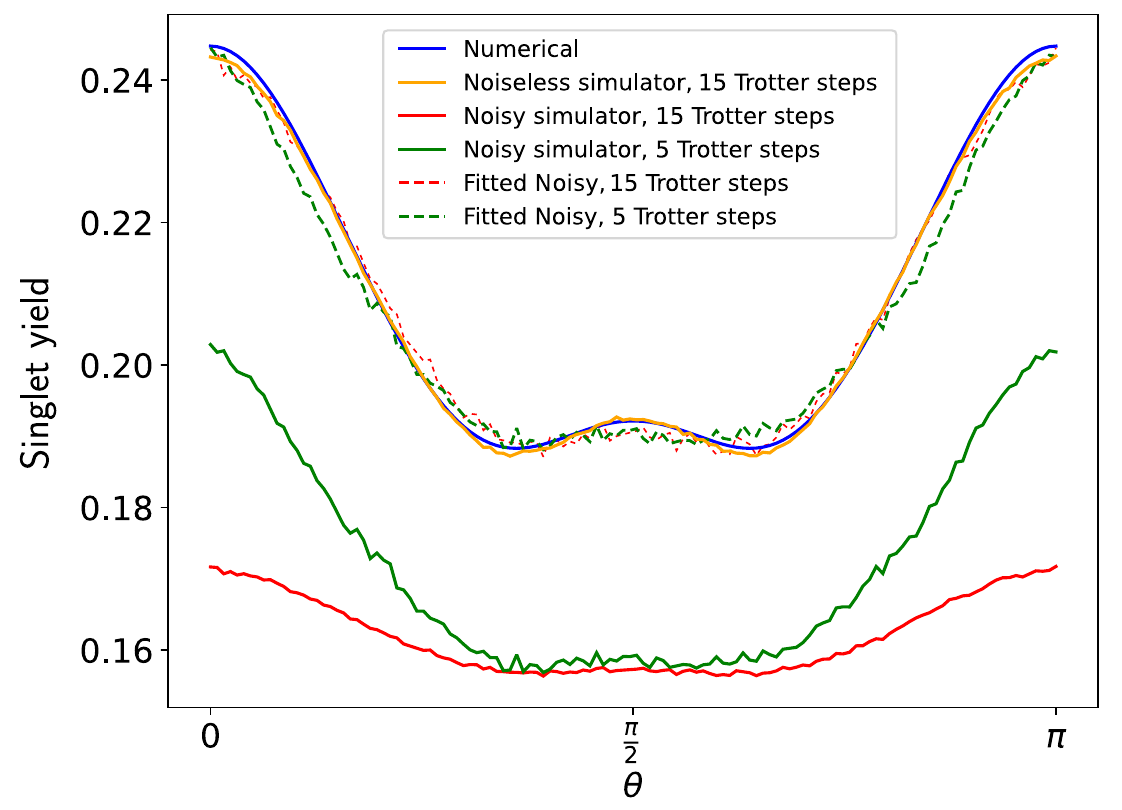}
	\caption{Singlet yield calculated with noisy simulator, noiseless simulator and numerical solution. Total evolution time was $1\mu s$ with timestep of $0.001\mu s$, the noisy simulator used the noise profile from \textsf{ibmq\_jakarta}. Noisy simulator results are observed to lose information resulting in a shift in singlet yield and decline in anisotropy. This is corrected by fitting to $\Phi_{S_{max}}$ and $\Phi_{S_{min}}$ of the numerical simulation, to assess the capability of reproducing the general dependence on $\theta$.}
	\label{fig_yields_renormal}
\end{figure}

By simulating the evolution of the singlet population for 128 angles (approximately 1.4 degree resolution in the range of $0 \leq \theta \leq \pi$), and evolving up to $1\mu s$,  we calculate the singlet yield as a function of $\theta$. The results from this calculation are shown in Fig. \ref{fig_yields_renormal}. It can be seen that noise introduces a shift in the singlet yield and diminishes the anisotropy, as quantified in Table. \ref{table1}. This behaviour is anticipated in accordance with previous studies of noise and spin relaxation in radical pairs simulated via master equations using classical open quantum systems simulations \cite{Vedral2011, Kattnig2016}. 
\begin{table}[!h]
	\begin{center}
		\caption{Anisotropy Values}
		\begin{tabular}{ |c|c| } 
			\hline
			Simulation & $\Delta_{S}$ \\
			\hline
			\hline
			Numerical & $0.0564$ \\
			\hline
			Noiseless & $0.0561$ \\
			\hline
			Noisy 15 steps & $0.0154$ \\
			\hline
			Noisy 5 steps & $0.0460$ \\ 
			\hline
		\end{tabular}
		\label{table1}
	\end{center}
\end{table}
To correct for this, we fit the noisy results to ensure agreement with the $\Phi_{S_{max}}$ and $\Phi_{S_{min}}$ values of the numerical prediction, thereby allowing an assessment of the simulators capability of reproducing the singlet yield behaviour with respect to the orientation to the magnetic field $\theta$. Results show reasonable agreement between the rescaled singlet yield behaviour as a function of $\theta$ predicted from numerical simulations, and with our method using the noiseless simulator. Whilst this agreement is lost with the introduction of noise, fitting to only the $\Phi_{S_{max}}$ and $\Phi_{S_{min}}$ values of the numerical simulation demonstrates that the overall reasonable agreement can be regained and thus, even in the noisy results, the singlet yield behaviour with respect to $\theta$ is retained.

\begin{figure}[t]
	\includegraphics[width=1.02\columnwidth]{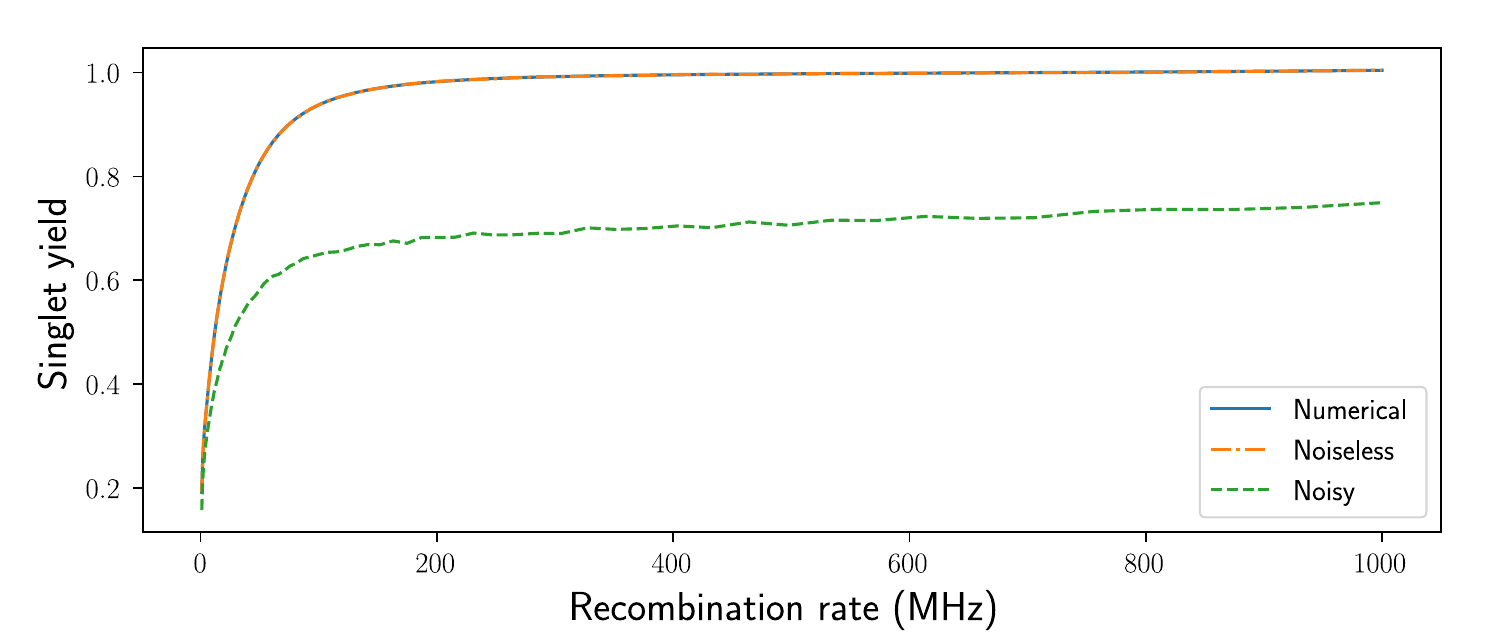}
	\caption{Singlet yield versus recombination rate for the same system, with $\theta=\pi/2$ and $n=5$. Both the numerical solution and noiseless simulation have similar behaviour. The noisy simulation exhibits a drop in singlet yield as observed when evaluating across $\theta$, but the curve has the same behaviour as the noiseless solutions.}
	\label{fig_recomrate}
\end{figure}

An additional check to understand if our system behaves as expected is shown in Fig.\ \ref{fig_recomrate}, where we compare how changing the recombination rate, $k$, changes the results from the singlet yield. The figure confirms how the noiseless simulation agrees with the numerical solution and shows the effects from noise on the final results. Similar profiles are observed for the noiseless and noisy simulations, with the noisy case providing better results as the recombination rate is increased. The noisy simulation has some deviation from the numerical and noiseless simulations, suggesting that for near term implementations, significant error is introduced even in the state preparation stage.
Despite this, as the fitting procedure in Fig. \ref{fig_yields_renormal} serves to illustrate, the general shape of the curve is conserved even in the presence of noise induced loss of amplitude, with the fit essentially mitigating for the loss. Oh et al.\cite{oh2023} shows a similar effect when simulating relaxation in the system. They simulate the radical pair dynamics in IBM QASM and add relaxation with the use of singular value decomposition. As the relaxation is increased, the angular dependency of the singlet yield decreases in a similar fashion to Fig. \ref{fig_yields_renormal}. Comparing the effect of simulated relaxation in Oh et al.\cite{oh2023} and the relaxation due to noise found in realistic quantum devices illustrates how one can use the noise as a tool in simulation algorithms.

\section{Challenges and Outlook}

The first application of a DQS scheme  to simulate the radical pair dynamics was due to Rost at al.\ \cite{rost}, where the simulation of thermal relaxation of radical pairs was attempted by decomposing the relaxation into Kraus operators and breaking it down into amplitude damping and dephasing channels. Subsequently, in our preceding work, we looked at a first attempt at incorporating hybrid quantum-classical machine learning techniques to adapt the model onto a simulation on an actual noisy quantum device \cite{trev_ieee}. While the use of the intrinsic noise in near-term quantum computers as a resource in simulations could prove to be a particularly fruitful strategy \cite{peetz2023simulation, jones23}, it is currently unknown if it would sufficiently emulate the noise arising from a biological environment for radical pairs associated with magnetoreception. Recent developments offer a potential resolution through manipulation and control of the intrinsic noise for desired decoherence rates of the open system dynamics \cite{Zhao2023, Guimaraes2023}. Here we have observed that there are similarities in the effects of noise and spin relaxation found in Fig. \ref{fig_yields_renormal}, that emerges from noisy simulator results using the noise profile of the \textsf{ibmq\_jakarta} quantum device, the results of Oh et al.\cite{oh2023}, which incorporate relaxation through a singular value decomposition, and previous master equation results simulated using existing open quantum systems simulation techniques \cite{Vedral2011}. This illustrates a possibly useful parallel in the increase of trotterization steps in our system and the increase of relaxation rates in QASM \cite{oh2023} simulations.

Another technique for simulating radical pair dynamics, proposed by Zhang et al.\cite{kais}, involves decomposing the non-unitary projection operators from the Lindblad master equation into Kraus operators. This approach mirrors the decomposition of thermal relaxation performed by Rost et al.\ \cite{rost} and Tolunay et al.\ \cite{jones23}, but instead of breaking the relaxation into damping and dephasing channels, they transpile the resulting matrix form of the time-evolution operator. Transpilation decomposes a matrix into a series of quantum gates, allowing for an approximation of the effect of the operator defined by the matrix on qubit states. Transpilation methods are known to result in increased circuit depths \cite{transp21}, thereby putting into question the relative efficiency of the method, and highlighting its lack of robustness to realistic device noise in the scaling limit for near term implementations \cite{Kim2023}. In contrast, our Trotter-Suzuki decomposition (Trotterization) method promises reasonable efficiency, for instance permitting an order of magnitude reduction in number of gates (on the order of 715 gates on noisy simulators), in comparison to Zhang et al.'s method requiring over 2000 gates to simulate a system of similar complexity. 

For the most part, our work conveys a sobering outlook on the potential of leveraging DQS towards realising radical pair spin dynamics simulations of utilitarian \cite{dWave, Kim2023, bubble} value beyond what is possible with state of the art classical emulation techniques in the immediate near term. Classical techniques have so far kept up with, and in many cases, exceeded, the concurrent advances in DQS tools \cite{cerezo2023, fastCS, orfi2024, hogg2024stochastic}, thus firmly positioning any possible prospects of achieving quantum advantage in a DQS context as a moving target \cite{schirber2024moving} as opposed to an inevitability. The key limitation of our method is the vulnerability to noise when increasing the number of Trotterization steps. But the crux of this is amenable to error mitigation techniques, further explorations of which are beyond the scope of our investigation at present. A primary objective for future studies incorporating error mitigation would be to tackle more hyperfine couplings by adding more qubits representing the nuclear spins to take advantage of the linear complexity scaling of computational resources in our DQS scheme, as opposed to the exponential scaling in existing classical techniques. Adding more nuclei would require the simulation of each state of the nuclear spins, or the inclusion of an auxiliary system. Adding more nuclei qubits also inevitably adds more noise, which we have found to be of significant impact even in simulations of the prototypical model presented in this work.  

Comparing to existing radical pair DQS schemes, we note that while Zhang et al.\ \cite{kais} propose a comparatively inefficient circuit construction which makes a practical near term implementation implausible due to bottlenecks in setting up transpilation efficiently in existing architectures and prohibitively long running times, they are able to formally incorporate non-unitary effects directly in the simulation. Tolunay et al.'s approach \cite{jones23} likewise allows for including non-unitary effects in the simulation, by using controlled noisy channels or controlling the inherent noise present in Noisy Intermediate-Scale Quantum (NISQ) devices. However, the primary noise sources considered in these earlier radical pair quantum simulations are the $T_{1}$ and $T_{2}$ thermal decays, which are not necessarily good surrogates for effects such as electron-electron dipolar interactions which are more typical of noise sources likely to be of relevance in magnetoreception. They also consider simulations of much shorter timescales, up to $60ns$, as opposed to the $1\mu s$ simulations presented in this work. 

Overall, we observe that noise in near term quantum devices can pose significant challenges in simulating radical pairs in existing quantum hardware, particularly in the context of emulating coherent spin dynamics via Trotterization. Furthermore, our finding that the singlet yield response to device noise is similar to noise arising from simulating the systems environment \cite{oh2023, Vedral2011}, could cause difficulty in distinguishing between these noise sources to accurately account for the open quantum system dynamics. This could be alleviated if the inherent device noise can formally be shown to reliably mimic the interaction of a biological environment and thus be utilized as suggested by Tolunay et al. \cite{jones23}, Beyond this, if the objective is to address the regime of radical pairs of biologically realistic complexity relevant to magnetoreception (which are beyond the scope of direct simulation using existing classical techniques) and the obstacle of quantum device noise can be overcome for at least the prototypical model thus far considered, then attention must turn to more advanced models. Specifically one must consider: increasing the number of hyperfine coupled nuclei using reliable parameters for FAD and tryptophan, including the unavoidable inter-radical interactions\cite{Babcock2020}, using asymmetric reaction kinetics, and accounting for realistic spin relaxation due to radical motion induced by the cryptochrome protein environment \cite{Kattnig2016}, all the while being conscious of the additional circuit complexity and consequent noise introduced. 

\section{Conclusions}
We implemented a Trotterization scheme to simulate the molecular and chemical reaction kinetics involving the singlet-triplet interconversion in the radical pair mechanism on a near term quantum simulator. Using this approach, we considered a simple model system consisting of two electrons and one spin-$\frac{1}{2}$ nucleus defined by a Hamiltonian comprising the Zeeman and hyperfine interactions, to calculate the singlet yields of the radical pair reaction. We identified the ideal Trotterization order for replicating known numerical results and demonstrated qualitative agreement in dynamics and in a more realistic noisy simulation, by replicating the response of the yield as a function of $\theta$, subject to scaling. 

There is an accuracy vs.\ noise resiliency trade-off in the choice for order of Trotter steps, as seen in earlier benchmarks of Trotterization in various contexts \cite{Lee_2023}. We have thus benchmarked how our method behaves subject to realistic device noise profiles, which required us to minimize the number of Trotterization steps. The main advantage of our method is efficiency in the scaling limit. It also lends itself for extension to include more nuclear interactions, allowing the simulation of more hyperfine interactions through the incorporation of more qubits. One could further use a Clebsh-Gordan expansion to represent particles with spin higher than one-half into a combination of qubits, for example, one can use two qubits to simulate a spin-1 particle. The method to define and later Trotterize the Hamiltonian, which we have presented, would be unchanged by these additions.

Future work could incorporate error mitigation techniques, and extend our method to directly include non-unitary effects during the quantum simulation, in the vein of Rost et al.\cite{rost}, where they attempted to harness the inherent noise of quantum devices to simulate thermal relaxation effects. Another next step would be to incorporate adaptive techniques for diminishing the effects of noise in Trotterization that are being developed \cite{trot24, Guimaraes2024, Zhao2023}. Overall, the method presented in this work enables the potential leveraging of near term quantum devices towards simulating the complex quantum dynamics of spin biological systems, and leaves open the promise of resolving near term simulation bottlenecks through efficient error mitigation techniques.

\begin{acknowledgments}
	PHA and MCO acknowledge support from the Coordenação de Aperfeiçoamento de Pessoal de Nível Superior – Brasil (CAPES) – Finance Code 001, National Council for Scientific and Technological Development (CNPq). PHA acknowledges support from the DERI/Santander 032/2022 International Mobility Fellowship towards a research visit and collaboration with the QuBiT Lab at UCLA, and the Deutsche Forschungsgemeinschaft (SFB 1372 Magnetoreception and Navigation in Vertebrates - Project Sig05). FTC and DRK acknowledge support from the Office of Naval Research (ONR Award Number N62909-21-1-2018). We acknowledge use of IBM Quantum services for this work. Views expressed are those of the authors, and do not reflect the official policy or position of IBM or the IBM Quantum team.
\end{acknowledgments}

\section*{Data availability}
The code and all scripts for reproducing the data presented in this paper are available from the corresponding author upon reasonable request.

\appendix

\section{Trotterization and Noise}\label{trotter_noise}
\begin{figure}[h]
	\centering
	\includegraphics[width=0.9\columnwidth]{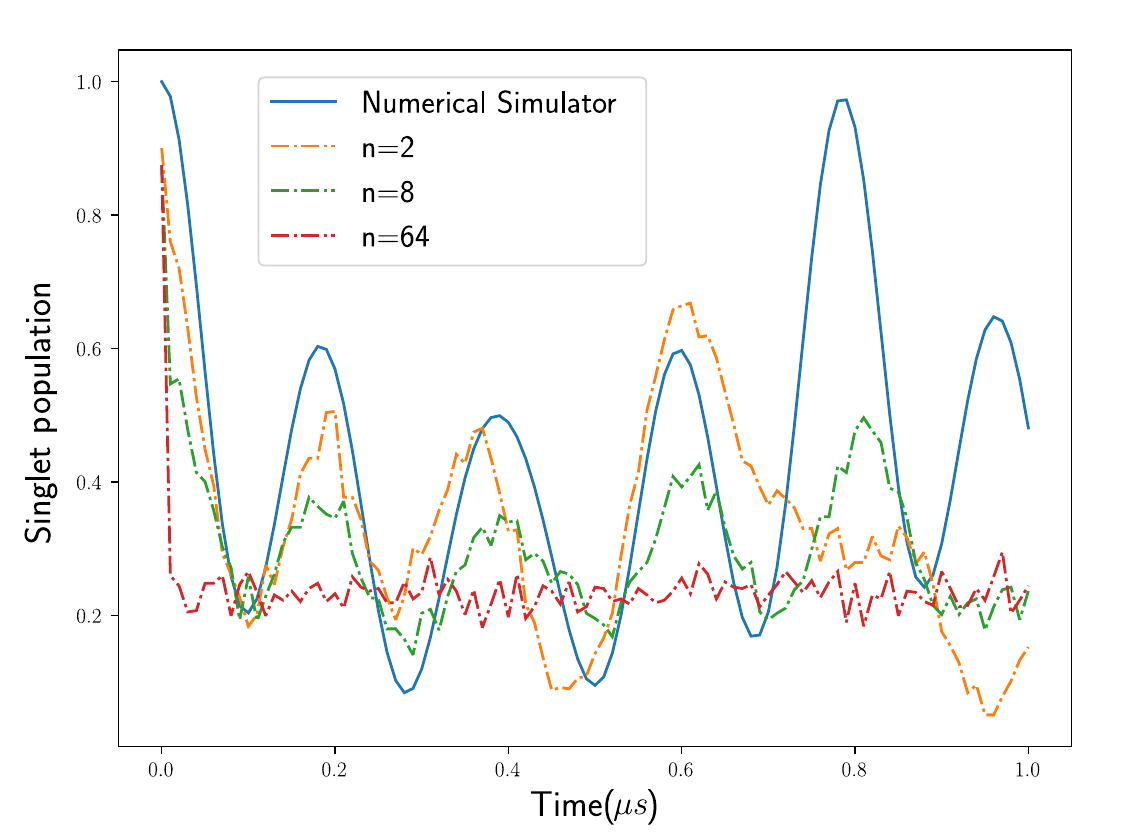}
	\vspace{-1em}
	\caption{Comparing the numerical solution for evolution of the population for $\theta = \pi/2$ with the method running on the noisy simulator for various orders of Trotterization $n$, again plotted up to $1\mu s$. This illustrates how increasing the order of Trotterization leaves the process more vulnerable to loss of amplitude to noise, we can see how the curves with less Trotterization steps maintain a higher amplitude and hold the shape of the curve better. We can compare with the noiseless curves from Figure \ref{fig_steps_noiseless}.}
	\vspace{-1em}
	\label{fig_steps_noisy}
\end{figure}

Existing quantum devices are not amenable to running circuits with inordinate gate counts, so we minimize the order of Trotterization, $n$, to tackle this challenge. 
\begin{figure}[h]
	\centering
	\includegraphics[width=0.855\columnwidth]{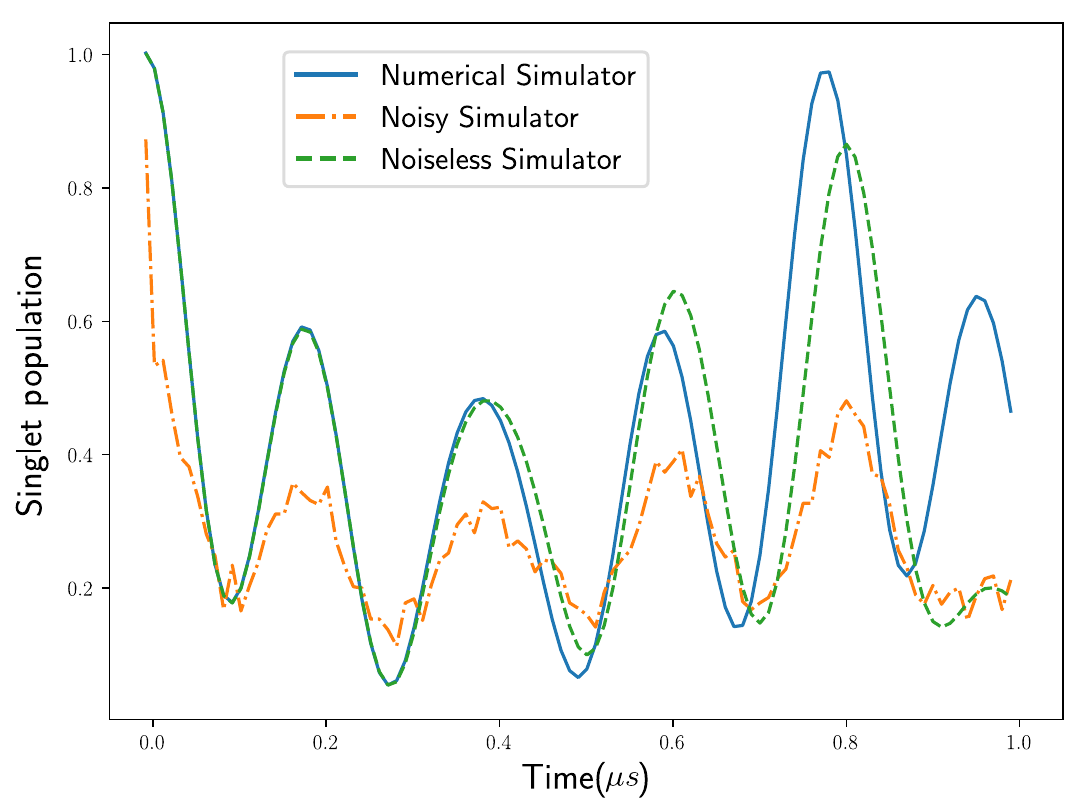}
	\vspace{-1em}
	\caption{Comparing the numerical solution for evolution of the population for $\theta = \pi/2$ with the method running on a noiseless simulator, and a noisy simulator with the same noise profile as IBM's \textsf{ibmq\_lagos} quantum device. The evolution is plotted up to $1\mu s$ to maximize the resolution while minimizing the number of samples that have to be acquired from the quantum device. To deal with the effects of noise and possible error during the application of a gate we also minimize the number of gates in the system by decreasing the number of Trotter steps from $1024$ in the previous simulations to $8$. This causes the noiseless simulation to drift from the numerical result, which is expected.}
	\label{fig_noisy}
	\vspace{-1em}
\end{figure}
One can see how increasing the order of Trotterization impacts the noise resilience of the circuit in Fig. \ref{fig_steps_noisy}.

This effect happens due to how Trotterization adds gates to the circuit, each gate has an error associated with it, so adding more gates also increases the probability of an error being introduced to the computation.  
\begin{figure}[t]
	\centering
	\includegraphics[width=0.85\columnwidth]{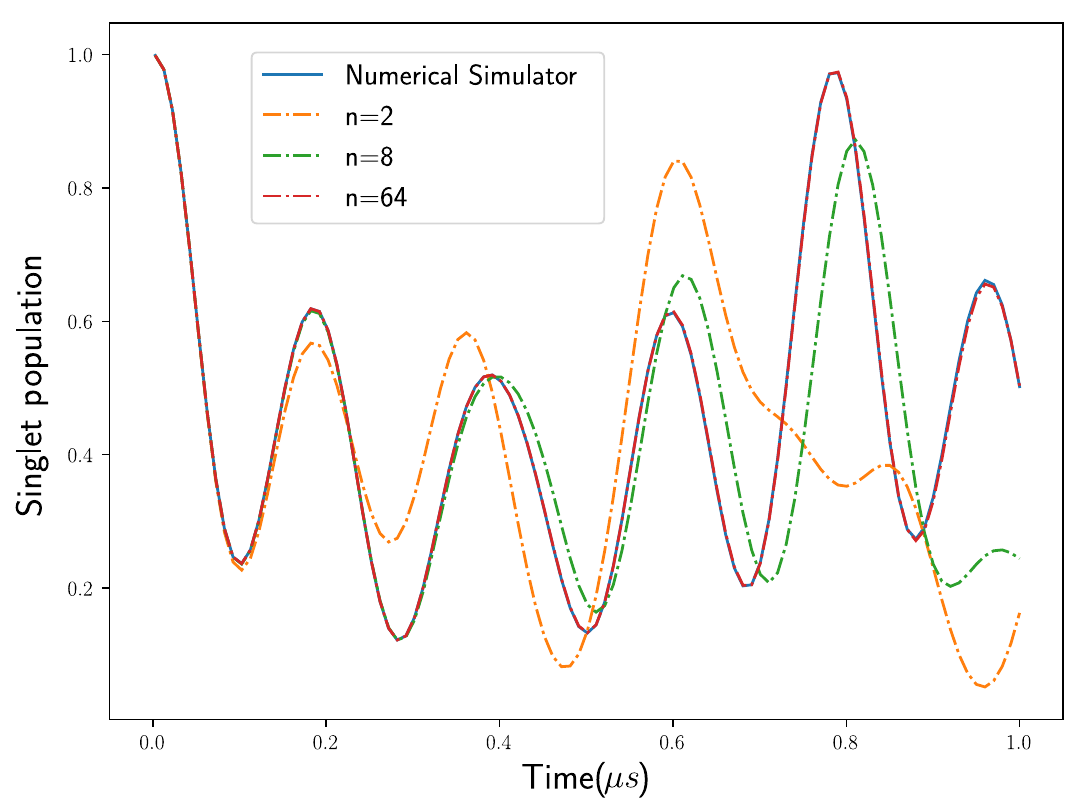}
	\vspace{-1em}
	\caption{Comparing the numerical solution for evolution of the population for $\theta = \pi/2$ with the method running on the noiseless simulator for various orders of Trotterization $n$. The evolution is plotted up to $1\mu s$ to maximize the resolution while minimizing the number of samples that have to be acquired from the quantum device. This illustrates how lower orders of Trotterization diverge earlier from the expected solution. This shows how Trotter-Suzuki is particularly fruitful for smaller evolution times, as it is very efficient and fairly noise resistant for shorter timescales (compare with Fig. \ref{fig_steps_noisy}), but on longer time scales we face the challenge of having to deal with a higher noise sensitivity for better approximations.}
	\vspace{-1em}
	\label{fig_steps_noiseless}
\end{figure}
One can better understand this by comparing the number of gates when using $n=1$ and $n=3$. For $n=1$, if we open all the gates in its fundamental components we have a total of 61 gates, where from those 55 are defined by the Trotterization. For $n=3$ the number grows to 171, where 165 are from Trotterization. This is a good sign, as the number of gates grows linearly with the order of Trotterization $n$, but as current quantum devices have difficulties running more than a few hundred gates without error correction we can see how this can become a problem, especially for larger systems. Therefore one should minimize the Trotterization order, but with a smaller $n$ the noiseless simulation drifts a bit from the numerical curve, as seen in Fig. \ref{fig_steps_noiseless}, which is to be expected from smaller orders of approximation. So there is the need to optimize the order of Trotterization given the presence of noise. 

\begin{figure} [h!]
	\centering
	\includegraphics[width=0.85\columnwidth]{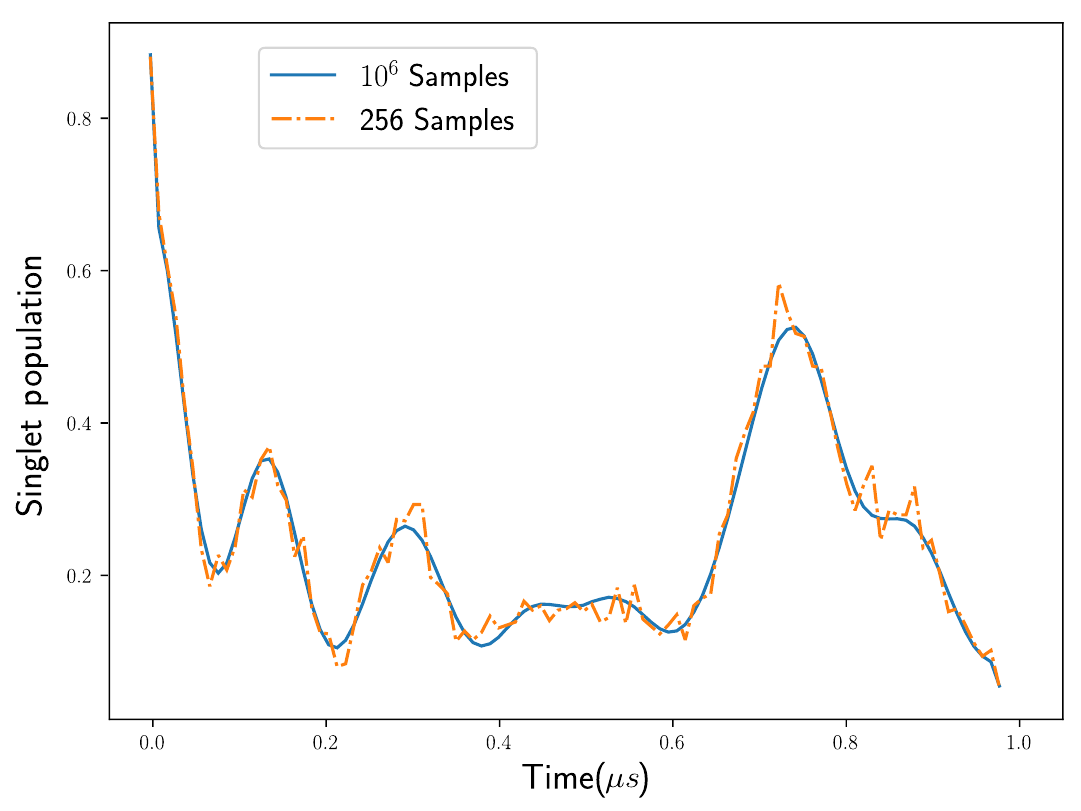}
	\caption{Singlet population curve relative to number of samples. Each sample corresponding to a single measurement taken at the end of each circuit run. As the number of samples is increased, the curve converges to a smooth shape.} 
	\label{fig_samplesize}
\end{figure}

Another important analysis is how the simulation improves as we increase the number of measurements. Each sample is one complete computation on the quantum computer, to get the proper statistics we need to increase the number of samples, but this also increases the computing time. In Fig. \ref{fig_samplesize} we show the effect of increasing the number of samples from the circuit. In general the circuit is behaving as expected, converging into a smooth curve as the number of samples increases.

\section{Time evolution \& rotation operator}

If we consider the mathematical form of the rotation operator \cite{Sakurai:1167961},
\begin{equation}\label{eq:def_rot_op}
	\hat{R}_x(\theta)\ket{\psi} = e^{\frac{-i\hat{X}\theta}{2}}\ket{\psi},
\end{equation}
and the unitary time evolution operator.
\begin{equation}\label{def_evo_op}
	\hat{U}(t)\ket{\psi} = e^{\frac{-i\hat{H}t}{\hbar}}\ket{\psi}.
\end{equation}
it is apparent that if the Hamiltonian can be written as a sum of Pauli operators then the unitary evolution of a closed system can be interpreted as a rotation, this applies to the Hamiltonian from Eq. \ref{eq:Hamiltonian}. The physical intuition for this come from the fact that Pauli operators cause the spin states to rotate. This can be observed easily in our system. The internal interactions of our systems all can be interpreted as effects causing the procession of spin states, or rotations around the Bloch sphere\cite{Bloch1946}. With this in mind, it is not difficult to see our Hamiltonian as a sum of one and two-qubit Pauli operators. If we open each spin operator from Eq. \ref{eq:Hamiltonian} we can better see how one would decompose a spin Hamiltonian into Pauli operators. 

$\hat{S}_X = \frac{\hbar}{2}\hat{X}$, where $\hat{X}$ is the Pauli-X operator, and the same applies to Pauli-Y and Pauli-Z. We can use the $\hat{\sigma}$ notation to simplify the Hamiltonian, where $\hat{\sigma}_1 = \hat{X}$, $\hat{\sigma}_2 = \hat{Y}$ and $\hat{\sigma}_3 = \hat{Z}$, this way we have
\begin{equation}\label{hamiltonian_decomposed}
	\begin{split}
		\hat{H} = \sum_{n=1}^2 g_n\mu_B\frac{\hbar}{2}(B_x\hat{\sigma}_{1n}+B_y\hat{\sigma}_{2n}+B_z\hat{\sigma}_{3n}) + \\ \sum_{i=1}^3\sum_{j=1}^3 \frac{\hbar^2}{4}a_{i,j}\hat{\sigma}_{i1}\hat{\sigma}_{jNucleus}.
	\end{split}
\end{equation}
Where $B_x$, $B_y$, and $B_z$ are the values for the external field in the respective coordinates, $a_{i,j}$ is the component in line $i$ and column $j$ of the matrix representation of the tensor $\bold{A}_{1,1}$, and $n$ is electron 1 or 2. Notice that as the nucleus is a proton with spin $\frac{1}{2}$, its spin operator $I_{1,1}$ is equal to the electron spin operator $\hat{S}_{1}$. We can then exponentiate the Hamiltonian to obtain the complete form of the time evolution operator
\begin{equation}\label{eq:evo_operator}
	\begin{split}
		\hat{U}(t) = e^{\frac{-it}{\hbar}\sum_{n=1}^2 g_n\mu_B\frac{\hbar}{2}(B_x\hat{\sigma}_{1n}+B_y\hat{\sigma}_{2n}+B_z\hat{\sigma}_{3n})} \\ e^{\frac{-it}{\hbar}\sum_{i=1}^3\sum_{j=1}^3 \frac{\hbar^2}{4}a_{i,j}\hat{\sigma}_{i1}\hat{\sigma}_{j1}},
	\end{split}
\end{equation}
which we can reorganize as 
\begin{equation}
	\begin{split}
		\hat{U}(t) = \Pi_{n=1}^2 e^{\frac{-itg_n\mu_B}{2}B_x\hat{\sigma}_{1n}}e^{\frac{-itg_n\mu_B}{2}B_y\hat{\sigma}_{2n}}e^{\frac{-itg_n\mu_B}{2}B_z\hat{\sigma}_{3n}} \\ e^{\frac{-it}{4\hbar}\sum_{i=1}^3 \sum_{j=1}^3 a_{i,j}\hat{\sigma}_{i1}\hat{\sigma}_{j1}}.
	\end{split}
\end{equation}

If we compare with Eq. \ref{eq:def_rot_op} we can see how we can rewrite its as a series of rotational operators
\begin{equation}\label{final_rot}
	\begin{split}
		\hat{U}(t) = \hat{R}_X(tg_n\mu_BB_x)\hat{R}_Y(tg_n\mu_BB_y)\hat{R}_Z(tg_n\mu_BB_y) \\ \sum_{i=1}^3 \sum_{j=1}^3 \hat{R}_{i,j}(\frac{t}{2}a_{i,j}),
	\end{split}
\end{equation}
where $\hat{R}_{i,j}$ represent simultaneous rotations in two spins, in this case the electron 1 and the nuclear spin. We use $i$ and $j$ to simplify the notation, where $\hat{R}_{1,1} = \hat{R}_{XX}$, $\hat{R}_{1,2} = \hat{R}_{XY}$, and so on. This example illustrates well the method to decompose the time evolution operator of a spin Hamiltonian into rotational operators. In Eq. \ref{eq:evo_operator} we intentionally opened the spin operator $\hat{S}$ into the three coordinates so that we have a final time evolution formed only by simpler rotations in those coordinates. This simplifies the calculation and makes it easier to visualize the final operator, it also provides a more convenient form to implement in the quantum computer, since those base rotations are fundamental gates. 

\section{Simulation parameters}\label{parameters}
Table \ref{table2} summarizes the parameters employed in the present simulations.
\begin{table}[!h]
	\centering
	\caption{Simulation Parameters}
	\begin{tabular}{ |c|c| }
		\hline
		$k_S$ & 1 MHz  \\ \hline
		$k_T$ & 1 MHz  \\ \hline
		$g$-factor & 2  \\ \hline
		$\mu_B$ & 57.8838 neV/mT  \\ \hline
		$B$ & 50 $\mu$T  \\ \hline
		$\textbf{A}_{1,1}$ & $\begin{bmatrix}
			5 & 0 & 0\\
			0 & 5 & 0\\
			0 & 0 & 10
		\end{bmatrix}$ neV\\ \hline
	\end{tabular}
	\label{table2}
\end{table}

\bibliographystyle{apsrev4-2}
\bibliography{refs}

\end{document}